\documentclass{jfm}
\usepackage{graphicx}
\usepackage{epstopdf, epsfig}

\usepackage{xcolor}
\usepackage{mathtools,mathrsfs, amssymb, bm}
\usepackage{hyperref}
\usepackage[capitalise]{cleveref}
\Crefname{figure}{figure}{figures} 
\usepackage{float} 

\usepackage{orcidlink}

\usepackage{fancyhdr}
\shorttitle{Sebastian et al., JFM 2024 }
\shortauthor{J. Sebastian, A. L. Schødt, and K. H. Jensen}

\title{Traversing a thin lubricant film in finite time}

\author{John Sebastian\aff{1},
  Alexander L. Schødt\aff{1}
 \and Kaare H. Jensen\aff{1} 
\corresp{\email{khjensen@fysik.dtu.dk}}
}
\affiliation{\aff{1}Department of Physics, Technical University of Denmark, Kongens Lyngby}

\begin{document}

\maketitle

\begin{abstract}

\noindent In this study, we investigate the dynamics of particles overcoming the hydrodynamic barrier posed by a thin fluid film to achieve contact in finite time, a phenomenon critical in various natural and engineered processes such as enzyme docking, catalysis, and vesicular transport. Using the framework of lubrication theory, which posits that drag force scales inversely with the film thickness, we explore how particles can achieve finite-time contact despite theoretical predictions of infinite time under constant force. We conduct experiments where a spherical particle settles under gravity and magnetic attraction, the latter introducing a spatially varying force. Our findings reveal that a spatially varying force significantly alters the settling trajectory, enabling finite-time contact. The results are supported by a simple model that links hydrodynamic drag and the impact of spatially varying forces. 
Finally, we illustrate that forces can be inferred from kinematic observations. 
In the future, this may provide insights into biological and microscale systems where direct force measurements are challenging. 
Our study demonstrates that varying forces can be harnessed to overcome lubrication barriers, offering potential applications in designing self-assembly systems and improving surface interaction processes.

\end{abstract}


\section{Introduction} \label{sec:intro}

Particle-wall and inter-particle interactions lie at the heart of numerous natural and artificial processes of wide applicability and relevance.
These include self-assembly of components \citep{cui_molecular_2023}, enzyme docking \citep{zhou_electrostatic_2018}, catalysis \citep{vogt_concept_2022}, and vesicular transport \citep{casadevall_vesicular_2009}, among others.
Each of these processes rely on particles approaching one another and achieving surface-to-surface contact in finite time.
However, since the vast majority of such processes are fluid mediated, the surfaces 
encounter a fluid film barrier that ought to be squeezed out before contact can be effectuated.
The hydrodynamic resistance, or drag, rendered by the thin film can be understood using the framework of lubrication theory,
which posits that the drag force scales inversely with the thickness of the fluid film.
Consequently, two ideal surfaces (i.e., void of surface roughness) approaching one another under the influence of a constant force 
can attain contact only in \textit{infinite} time \citep{stone_lubrication_2005-3}.

It is despite this impediment that we observe particles and particle analogues achieving contact in finite time
in a wide variety of scenarios spanning natural and engineered realms. 
In fact, in biological processes,  the sustenance of biochemical processes (e.g., enzyme docking) often rely on particles overcoming the separation 
due to the surrounding solvent.
More importantly, the process of achieving `first contact' has to happen in finite time - at a scale relevant to the reaction rate, $~10^{-6} - 10^{0}$s \citep{agarwal_enzymes_2006}.
Another instance where this is relevant is in the process of cell wall deposition in plant and fungal cells \citep{verma_cytokinesis_2001,casadevall_vesicular_2009}.
The process relies on the timely deposition of cellulose matrix onto a newly formed cell barrier by individual spherical vesicles within the time constraints
imposed by the process of cell division \citep{thiele_timely_2009}. 
Similarly, at smaller spatial scales, we have the adsorption of phoretic particles (intracellular macromolecules, colloids, etc.), onto channel walls and onto one another.
Once again, the relevant binding processes tend to manifest at rates far higher than those expected for such particles, 
given the existence of a thin film of fluid at the interface \citep{leckband_intermolecular_2001,xu_dynamic_2023}.

While the examples invoked here highlight the importance of the oft neglected contribution of hydrodynamic forces that
significantly alter the expected path and interactions between particles (as in \citet{squires_like-charge_2000-1}, for instance), perhaps more importantly,
they point towards the key physical mechanism that may be employed to cross over the thin film barrier in finite time, and/or at faster rates.
Specifically, motivated by the numerous instances presented above, and others we will discuss later in \cref{sec:varying},
the predominant scheme that allows finite-time contact appears to be the utilisation of a spatially varying force.
Depending on the origin and nature of the interaction, the spatial dependence of such a non-constant force may follow an arbitrary law. 
Nonetheless, for a variety of cases, where the fundamental nature of the interactions involved are known, the effective force can be evaluated for specific geometries. 

\begin{figure}
    \centerline{\includegraphics[width=1\linewidth]{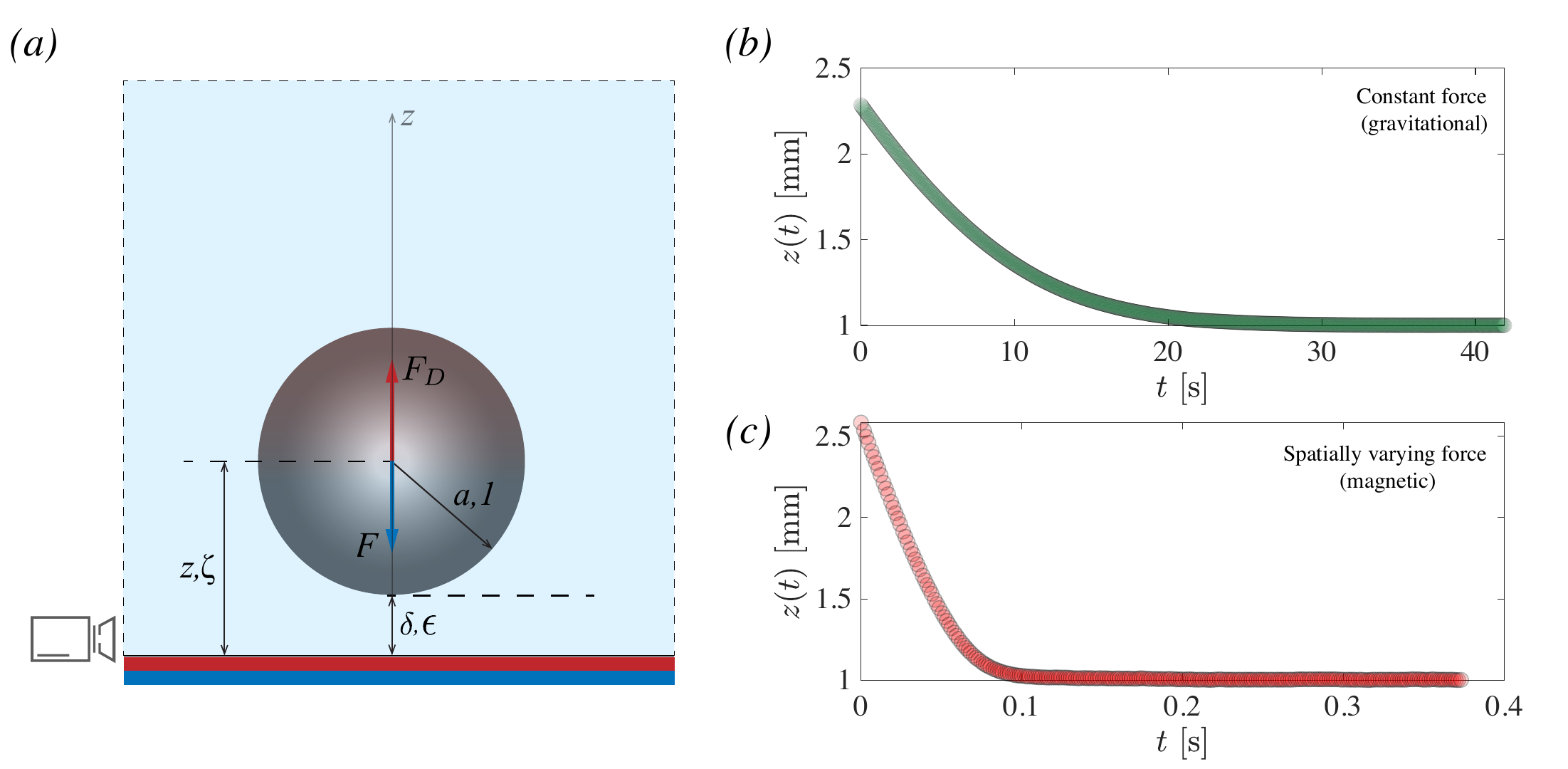}}
    \caption{(a) Schematic of the settling experiment system. 
    A metallic sphere of radius $a$ approaches a flat plane at the base driven by a force $F$ and subject to hydrodynamic drag $F_D$ in column of viscous silicone oil.
    The instantaneous center of the sphere is denoted by $z$, (nondimensional position, $\zeta = z/a$) and the corresponding separation between the sphere and plane, i.e., smallest gap is given by $\delta$, (nondimensional gap, $\epsilon = \delta/a$).
    Experimentally determined trajectories of a sphere of radius $a = 1 \ \mathrm{mm}$ - (b) settling under gravity, i.e., a constant force, and 
    (c) driven by magnetic attraction to the stationary permanent magnet placed at the base.}
     \label{fig:fig1}
\end{figure}

Therefore, to explore the relevance and consequences of this strategy, we need to make use of a spatially varying force that is both experimentally and theoretically tractable. 
To make an initial advance in this direction, we invoke the classical system of a spherical particle approaching a rigid plane (a wall or floor) in a quiescent fluid, as illustrated in \cref{fig:fig1}(a) (see \cref{appA} for details of the experimental setup and methods). 
Here, a sphere of radius, $a$, approaches a flat plane in a semi infinite fluid domain, driven by a force, $F$.
At any given instant, the sphere's center is at a height, $z$, above the floor plane, which is fixed at $z=0$.
To illustrate this process, we consider the simple experimental system where a  steel sphere (radius $a=1$ mm) settles under its own weight, i.e., $F =\mathrm{const.}$, in a cuboidal container of silicone oil. 
The sphere's unidirectional, downward trajectory is observed orthogonally using a camera, 
focussing on the region close to the floor (interface) to capture its final path as it overcomes lubrication stresses. As expected, the initial trajectory is linear, 
and the terminal settling velocity is consistent with Stokes' drag formula (\cref{fig:fig1}(b)). 
Closer to the surface, i.e., when the sphere's position is comparable to the particle radius, $z\approx a$, drag increases, and the vertical speed decays. 
After a few tens of seconds, movement is no longer detected.

In order to test and verify the strategy of using a spatially varying force, to accelerate this process, we introduce a magnetic modification to the settling experiment described above. 
By simply introducing a commercially available permanent magnet (\cref{appA}) at the base of the fluid container for the sphere to settle onto, we introduce a spatially decaying magnetic field in the fluid domain.
In turn, the metal sphere behaves as an induced spherical magnet, effecting a net attractive force, $F$, that depends on $z$, and is strongest at the floor plane. 
As with the gravity-driven settling experiments, we track the trajectory of the sphere as it approaches the base magnet. 
We note two major differences between the gravitational (\cref{fig:fig1}(b)) and magnetic cases (\cref{fig:fig1}(c)): first, the induced magnet reaches the surface approximately a hundred times faster. 
(The absolute time saved depends, of course, on the magnet's strength, discussed in \cref{sec:magnetic}).
Second, it appears that the spatially varying force has extended the linear regime, i.e., the Stokes-like motion. 
A clear deviation from the linear trajectory is thus only observed when the particle-wall separation $\delta \approx 0.1 a$ (nondimensional gap, $\epsilon \approx 0.1$), is comparatively small. 

To explore the origins of these quantitative and qualitative effects, we conducted the experiment described above using spheres of eight different diameters $2a = 1, 2, 2.5, 3.2, 3.5, 4, 4.5, 5$ mm, 
tracking their end trajectories when driven by gravitional and magnetic forces.
In the following sections, we will first develop the theoretical framework (\cref{sec:curse_lub}) and conduct a formal analysis of the experimental data. 
This entails a description of the hydrodynamic drag, its mathematical form, and its implications in \cref{sec:drag}.
We will see that the trajectory of a sphere approaching a plane under the influence of a general force, $F$, whose spatial dependence is captured by a function, $f(\epsilon)$, such that $F \sim f(\epsilon)$, follows
\begin{equation}
        \frac{\dot{\epsilon}}{\epsilon} + f(\epsilon) = 0,
    \label{eq:norm_dyn}
\end{equation}
which governs the evolution of the nondimensional gap, $\epsilon$, and serves as a master equation for subsequent analyses.
In \cref{sec:sediment}, we will look at the case of gravity-driven settling, which corresponds to $f(\epsilon) = 1$ in \cref{eq:norm_dyn}, which also serves as a first validation of the theoretical model.
This is followed by a discussion on the consequences of a spatially varying (magnetic) driving force in \cref{sec:varying}.
Further, we will utilise the outcomes to show that one can determine the nature of unknown spatially varying forces from kinematic observations in \cref{sec:estimate_forces}.
The benefits of such analyses are two fold -- one, a wide variety of binding forces relevant to biology and microscale systems which are inaccessible for direct force measurements, can be determined from non-invasive kinematic observations; 
two, the results can conversely be used to design self-assembly systems that utilise specific types of inter-particle interactions to overcome the lubrication barrier, to achieve desired structures. 
In \cref{sec:snap}, we will then look at other varying forces whose spatial dependence is known and explore their outcomes, followed by a brief discussion in \cref{sec:conclusions}.

\section{Theoretical framework} \label{sec:curse_lub}

Our system, introduced in the previous section, consists of a cuboidal reservoir of highly viscous fluid,
in which a metallic sphere of radius, $a$, travels in pure translation (i.e., without rotation) vertically downward along its centre, as depicted in \cref{fig:fig1}(a). 
The fluid used in our experimental setup is a silicone oil of density $\rho_f = 980 \ \mathrm{kgm^{-3}}$ 
and dynamic viscosity $\mu = 59 \ \mathrm{Pa s}$ ($\thickapprox 6 \times 10^4$ times more viscous than water).
The coordinate of the center of the sphere is $z$ at any instant, $t$.
The instantaneous gap between the two surfaces, i.e., the floor plane (at $z = 0$) and the base of the sphere is therefore $\delta = z - a$.
Normalising all length scales in the system by the radius, $a$, we have the non-dimensional centre coordinate $\zeta = z/a$ and 
non-dimensional gap thickness $\epsilon = \delta/a$, related by $\zeta = 1+\epsilon$. 

We will use the above coordinate system to write the equation governing the motion of the sphere.
The total force experienced by the immersed particle is the sum of net body forces and surface stresses acting on it
\begin{equation}
    \bm{F_{\mathrm{total}}} = \int \limits_V \bm{f_b} \ \mathrm{d}V + \int \limits_S \bm{\sigma \cdot n} \ \mathrm{d}S
    \label{eq:total_force}
\end{equation}
where $\bm{f_b}$ is an arbitrary body force (such as gravity or magentic attraction, see \cref{fig:fig1}) acting over the volume $V$ of the particle, and $\bm{\sigma}$ 
denotes surface stresses over $S$, the surface area of the particle. The local normal vector on $\mathrm{d}S$ is denoted by $\bm{n}$.
The surface stresses acting over the bounds of the particle may be of myriad origins. 
For example, for dielectric particles in an external electric field, $\bm{\sigma}$ will constitute Maxwell stresses \citep{wang_particle-surface_2022}, 
and for colloids in an electrolytic medium, it will include solute-surface interactions \citep{michelin_phoretic_2014}. 

Regardless of the specific interactions dominant in a given scenario, particles in all fluid systems will encounter hydrodynamic stresses.
Hence, it is prudent to separate the total force, $\bm{F_{\mathrm{total}}}$, into two components, 
the total hydrodynamic drag, $\bm{F_D}$, and the net external driving force, $\bm{F}$. 
As the particle motion is unidirectional along a vertical, rectilinear path in our settling experiments,
vector notation can be dropped.
Therefore, we can simply write the force balance along the vertical $z$ axis as $F_{\mathrm{total}} = F - F_D$.
Adopting this simpler notation allows us to resolve the hydrodynamics problem separately from that of the driving force, $F$.
We can immediately write, for the case of gravitational settling, $F = F_g = \mathrm{const.}$
Similarly, a spatially varying force, such as the induced magnetic attraction in our system, can be written as
\begin{equation}
    F = F_0 f(\epsilon).
    \label{eq:gen_pow_law}
\end{equation}
Here, $F_0$ is a constant (with dimensions of force) and $f(\epsilon)$ is a function of the non-dimensional gap, $\epsilon$. 
Following this general notation, $F_g$ is a special instance of the generalised force $F$ for $f(\epsilon) = 1$.

Irrespective of the nature of the driving force, the sphere is subject to hydrodynamic drag, $F_D$, which depends on its proximity to the wall.
Hence the effective trajectory of the particle is determined by the interplay between the driving force and fluid drag.
\begin{equation}
    \rho_s V  z^{\prime \prime} = F - F_D,
    \label{eq:force_balance}
\end{equation}
where time derivatives are denoted by $()^\prime$.

Since the left-hand side of the above equation represents the inertia of the particle,
we can immediately examine its relevance in determining its trajectory.
To do so, we can estimate the starting/stopping time for inertial motion of the particle as the ratio of the 
particle's momentum and the hydrodynamics forces retarding its motion as $\rho_s V / \mu a \sim 10^{-4}$s.
In contrast the kinematics of the settling trajectories we study span $\sim 10^{-1} - 10^2$s.  
Similarly, any rotational motion of the sphere in the experiment is also \textit{instantaneously} brought to rest.
In addition, we can inspect the typical Reynolds number, $\Rey = \rho_f u a/ \mu$, characteristic to the fluid motion induced by the translating sphere.
In fact, the choice of fluid in our experiments was inspired by the numerous examples of inter-particle and surface-particle interactions invoked in the previous section 
which share the feature that particle kinematics are \textit{slow} and restricts the relevant Reynolds number,  to $\Rey \ll 1$.
The maximum $\Rey$ encountered in our experiments is $ \approx 5 \times 10^{-6}$.
As a result, we can ignore the inertial term, simplifying the force balance in \cref{eq:force_balance} further, as we will see in the next section.
Hence, what we now require to determine the trajectory, $z(t)$ of the particle, are the expressions for the driving force, $F$, 
or more specifically, the function $f(\epsilon)$, and that of the hydrodynamic drag, $F_D$.  
To this end, we will first look at the drag force experienced by a sphere that is approaching a flat plane.

\subsection{Hydrodynamic drag} \label{sec:drag} 
Leveraging the low-Reynolds-number dynamics of fluid motion, the quasi-steady particle trajectory can be deduced to simplify \cref{eq:force_balance} and follow the force balance $F = F_{D}$.
Furthermore, as we are interested in the dynamics close to contact, we ought only to lay focus on the thin gap limit, i.e., $\epsilon \rightarrow 0$.
Nonetheless, we begin by briefly exploring how hydrodynamic drag evolves as a particle approaches a plane, 
from an initially unbounded region (`sky') to the thin film limit (`floor'), which is explicitly manifested in a sedimentation system such as ours. 
This corresponds to the early time progression of the trajectories shown in \cref{fig:fig1}(b) and (c).
We first consider the free regime where the particle moves in an unrestricted, infinite fluid domain whose walls impose no effect on the particle's motion.
The hydrodynamic drag on a sphere moving in such an unbounded fluid domain is given by the well-established Stokes' law of resistance, 
which captures its linear dependence on its instantaneous velocity, $u$, and radius, $a$.
As the sphere approaches the floor, the velocity and pressure fields are modified by the proximal presence of the planar surface, 
both fields no longer decay symmetrically away from the particle \citep{happel_low_1983-1}. 
This results in an increase in drag which reflects in its trajectory as the notable deceleration in \cref{fig:fig1}(b) and (c).

Thus, at least three distinct drag regimes can be identified from observing the path of a settling sphere. 
As depicted in \cref{fig:fig1}(b) and (c), far away from the floor, i.e., $\epsilon \rightarrow \infty$ or equivalently $\zeta  \gg 1$, the sphere progresses at a constant velocity. 
This is the unbounded or Stokes' regime, where the hydrodynamic drag experienced by the sphere is given by 
$F_{\text{Stokes}} = - 6 \pi \mu a u$. The negative sign captures the direction of the force on the particle with respect to $u$.
As the sphere approaches the wall, the hydrodynamic drag increases, 
and transitions from $F_{\text{Stokes}}$ to $F_D  = F_{\text{Stokes}}(1 + \frac{9}{8} \frac{1}{\zeta} ) $. 
This may be referred to as the Lorentz regime \citep{lorentz1897general,hasimoto_lorentzs_1996}. Note that, here, the additional drag varies inversely with the instantaneous 
coordinate of the sphere center, $\zeta = 1 + \epsilon$. This regime is relevant where $\zeta \gtrsim 10$ but finite (see §7.4 of \citep{happel_low_1983-1}).  

Finally towards the end of its trajectory, as the sphere and the floor is separated only by a relatively thin film of fluid between them, 
i.e., as $\epsilon \rightarrow 0$, the hydrodynamic drag is dominated by lubrication pressures. 
At this limit, the drag force converges to the form 
\begin{equation}
    F_{D} = F_{\text{Stokes}} \left[ \frac{1}{\epsilon} + \frac{1}{5}\ln{\frac{1}{\epsilon}} + k  \right],
    \label{eq:cox_lub}
\end{equation}
where $k = 0.971264$ is a constant that recovers numerical accuracy with the full series solution derived by \citep{cox_slow_1967}.

Having estabilshed a relatively complete physical picture of the approach, we once again focus on the experimental system (\cref{fig:fig1}), i.e., the final trajectory of the sphere as it approaches the flat plane. I this limit, we can further simplify \cref{eq:cox_lub} by 
retaining only the dominant term, to recover the familiar lubrication drag law, attributed to G. I. Taylor \citep{cox_slow_1967}
\begin{equation}
    F_D = F_{\text{Stokes}}  \frac{1}{\epsilon} =  - 6 \pi \mu a^2 \frac{\epsilon^\prime}{\epsilon}.
    \label{eq:lub_drag}
\end{equation}
This asymptotic simplifiation deviates from the exact solution, Eq. 2.2 of \cite{cox_slow_1967}, only by $\sim 3 \%$ even as $\epsilon \sim 1/2$ and is exact as $\epsilon \rightarrow 0$. 
(As we will see in later sections, our experimental data also falls within this range of $\epsilon$.)

The notable feature of the drag (\cref{eq:lub_drag}) at this limit is that $F_D$ varies inversely with the non-dimensional thin film gap $\epsilon$.
Consequently, the dynamic law governing the trajectory of the sphere, \cref{eq:force_balance} becomes
\begin{equation}
    6 \pi \mu a^2 \frac{\epsilon^\prime}{\epsilon} + F_0 f(\epsilon) = 0.
    \label{eq:gen_dyn}
\end{equation}
Finally, simplifying this leads to the non-dimensional equation for the temporal evolution of the non-dimensional gap, $\epsilon$, we recover \cref{eq:norm_dyn},
        ${\dot{\epsilon}}/{\epsilon} + f(\epsilon) = 0$, 
where the time scale that emerges, $T = {6 \pi \mu a^2}/{F_0}$, is used to normalise time, giving the non-dimensional time,
$\tau = t/T$. Henceforth, differentiation by $\tau$ will be denoted by $\dot{()}$.

Now that we have simplified \cref{eq:force_balance} to \cref{eq:norm_dyn} at the thin gap limit, 
we are ready to probe the effects of various spatially varying forces (i.e., arbitrary $f(\epsilon)$) on the end trajectory of a sphere approaching a plane. But before commencing that exercise, we will substantiate the methodology using the familiar case of gravity driven sedimentation ($f(\epsilon) = 1$). 
    
\subsection{Gravity driven sedimentation} \label{sec:sediment}

\begin{figure}
    \centerline{\includegraphics[width=0.75\linewidth]{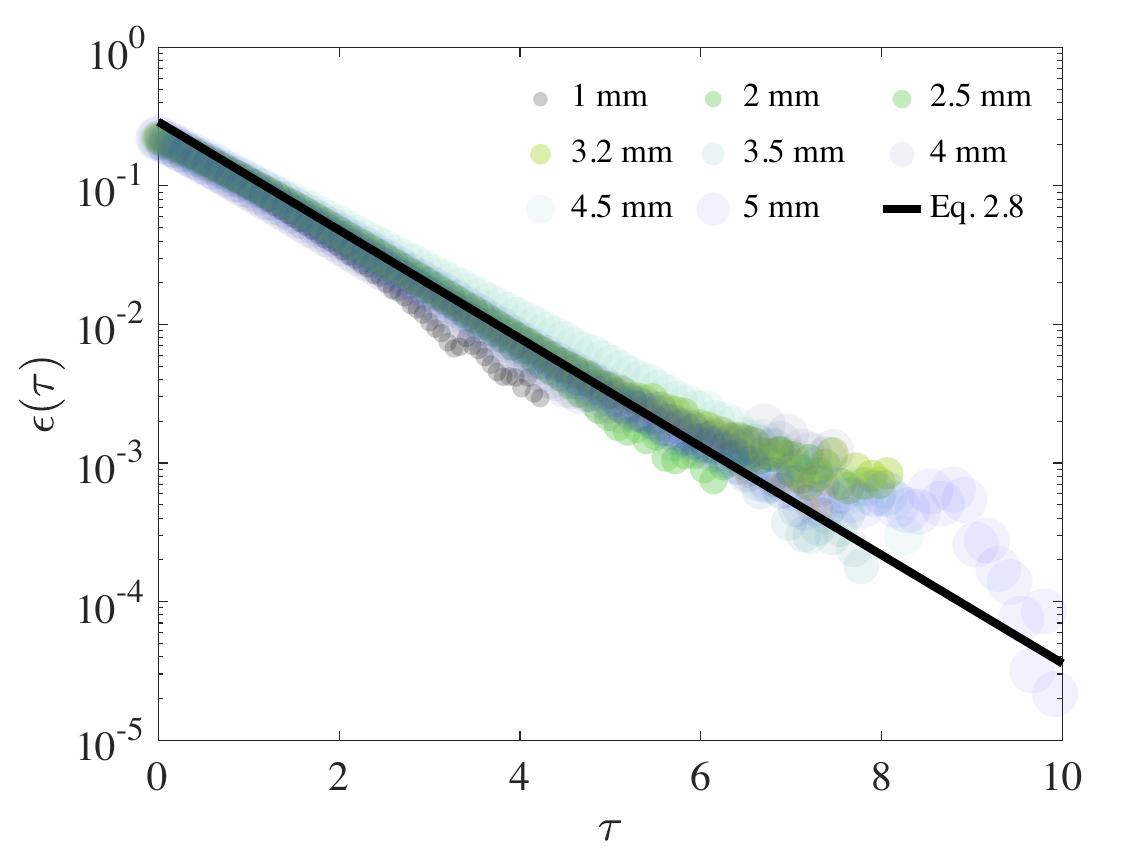}}
    \caption{End trajectories \(\epsilon - \tau\) of spheres different radii, $a$, settling under gravity.
        The expected exponential trajectory,\cref{eq:grav_sol}, is the black line.}
    \label{fig:grav_tracks1}
\end{figure}

To gauge the validity of \cref{eq:norm_dyn}, we will begin by considering the case of the sphere settling under gravity, where $f(\epsilon) = 1$,
and compare it against the trajectory data, $\epsilon - \tau$, presented in \cref{fig:grav_tracks1}.
In this case, the constant driving force, $F$, is the effective weight of the particle, $F_g  = \rho_s V \left( 1 - \sigma \right) g$, 
where $\rho_s V $ is the mass of the sphere of volume $V \sim a^3$ and density $\rho_s$, and $g$ is accelaration due to gravity.
The density of the fluid, $\rho_f$, is normalised by the density of the particle, $\rho_s$, to give $ \sigma = \rho_f / \rho_s $.
Since the particle is heavier than the surrounding fluid, $\sigma \ll 1$. 
In \cref{eq:norm_dyn}, along with $f(\epsilon) = 1$, we therefore have $F_0 = F_g$ and $T = {9 \mu}/{2 \rho_s (1 - \sigma) a g}$, giving 

\begin{align}
    & \frac{\dot{\epsilon}}{\epsilon} + 1  = 0 \label{eq:grav_eqn}\\
    \implies & \epsilon(\tau) = \epsilon_0 e^{ -(\tau - \tau_0) } 
    \label{eq:grav_sol}   
\end{align}
where $\epsilon_0 = \epsilon(\tau_0)$ is the initial condition, where $\tau_0$ represents the time beyond which the above asymptotic solution holds.
This exponential evolution of gap closure under a constant force such as gravity takes us back to our initial discussion about how 
the efficiency of thin film lubrication manifests unintentionally as a primary barrier to various natural, biological, and artificial processes that requires interparticle and 
particle-wall contact to occur in finite time. 
The concurrence of experimental data with the solution \cref{eq:grav_sol} is evident in \cref{fig:grav_tracks1}, 
where the experimental data, with length and time scales are respectively normalised by $a$ and $T$ are plotted.
The black line is \cref{eq:grav_sol}, where the initial condition, $\tau_0$,  is chosen such that $\epsilon_0 \rightarrow 1/2 $,
per the condition for the validity and accuracy of this limiting solution, that was discussed in \cref{sec:drag}, (below \cref{eq:lub_drag}).

Here, the expression for the sedimentation time scale, $T \sim (\rho_s a)^{-1}$, suggests that a larger, heavier particle will fall at a smaller time scale of the 
exponential decay, even though gap closure would still occur only as $\tau \rightarrow \infty$ no matter what the magnitude of the constant force is. 
This well-studied result \citep{mongruel_approach_2010,xu_dynamic_2023} that we have also revalidated experimentally underlies the familiar problem of `infinite contact time' in interparticle adhesion \citep{israelachvili_intermolecular_2011}.

It is in this context that the strategy of using a non-constant, spatially varying driving force becomes relevant.
In fact, systems in the natural world, where particles or surfaces approach each other in a fluid medium, are rarely driven by constant forces \citep{brangbour_force-velocity_2011}.
In the following section, we will explore how forces that follow a non-trivial law, $f(\epsilon)$, modify the kinematics of sedimentation.

\section{Particles driven by a spatially varying force} \label{sec:varying}

Forces between particles and surfaces are seldom constant, 
especially at progressively smaller spatial scales. 
This is quickly recognisable in colloidal systems where the forces between surfaces depend on their surface charges \citep{squires_making_2008}.
More broadly, forces between particles, particle analogues, surfaces, and interfaces are often cumulative forces originating from their material properties and those of the surrounding fluid.

The key feature of these forces, regardless of their physical origin in each case, is that the effective driving force is a function of the 
thickness of the instantaneous gap separating the surfaces. 
A brief survey of such forces relevant to surfaces in close proximity demonstate that a number of them typically follow power-laws of the form $f(\epsilon) = \epsilon^{-n}$, or $f(\epsilon) = (d_0+\epsilon)^{-n}$, 
where $d_0$ is a (non-dimensional) length scale and $n$ is an exponent.  
In \cref{force_table}, we catalogue a selection of interactions that exhibit power-law behaviour of either form.
The exponent, $n$, in $f(\epsilon)$ is the primary factor that dictates the kinematics of the particle,
and the values or ranges of $n$ for a spherical particle driven towards a plane are tabulated.
Once again, we retrieve for $n=0$ gravity (and other constant forces).
Another notable example in \cref{force_table} is the van der Waals adhesion between a sphere and a plane, which follows the power-law, $f(\epsilon) = \epsilon^{-n}$, where $n=2$. (For a selection of force laws for van der Waals interactions between other geometries, refer Sec. 13.2 of \citet{israelachvili_intermolecular_2011}. 
Note that $n$ depends on the geometry of the interacting surfaces. While we have limited our analysis to the sphere-plane geometry in this work, it can be extended to particle pairs of other geometries).

\begin{table}
    \begin{center}
\def~{\hphantom{5}}
        \begin{tabular}{ lcc } 
            Interaction & Scaling law, $f\left(\epsilon\right)$ & Exponent, $n$ \\ 
            Gravity                                     & $\epsilon^0$              & 0\\ 
            Stefan adhesion                             &  $\epsilon^{-3}$          & 3\\
            van der Waals                               &  $\epsilon^{-2} $         & 2\\
            London                                      &  $\epsilon^{-7} $         & 7\\
            Charged Colloid – charged surface           &  $\epsilon^{-n} $         & 0 - 1 \\
            Ion - Ion; Ion-Dipole; Dipole-Dipole        & $( d_0 + \epsilon)^{-n} $ & 0 - 4 \\ 
            Metal sphere-Permanent magnet (this study)  & $( 1 + \epsilon)^{-n}$    & 0 - 4\\
        \end{tabular}
    \end{center}
\caption{A condensed list of known interactions of various origins. Such spatially varying forces typically exhibit a power-law relation with the gap $\epsilon$,
 of the form $f(\epsilon) = \epsilon^{-n}$ or $f(\epsilon) = (d_0 + \epsilon)^{-n}$, where $d_0$ is a length scale  inherent to the system. 
The numerical range into which the value of the exponent, $n$, for each force type is listed. For well-known named interactions, $n$ is determined from the corresponding
interaction energy, $\mathit{U}$, as $F = -\bm{\nabla} \mathit{U}$. 
For the other interactions, the value of $n$ depends on other system parameters, and its range is summarised from prominent literature \citep{israelachvili_intermolecular_2011,leckband_intermolecular_2001}. }
\label{force_table}
\end{table}

On the other hand, the magnetic interaction implemented in our experimental system follows the second type of power-law, namely, $f(\epsilon) = (d_0+\epsilon)^{-n}$, where $d_0\approx 1$ and $n\approx 0.75$. 
This scaling is based on independent measurements of the induced magnetic force (see \cref{appB}).
The specific functional form can be understood from a brief consideration of classical electromagnetic/ electrostatic interactions between point charges, dipoles, etc., 
whose dependence on the separation are well understood \citep{smythe_static_1989-1}, of which the inverse square law for point charges is perhaps the most familiar example.
In these cases, the spatial dependence is typically a function of the centre-to-centre distance, i.e., $ \epsilon + d_0$, 
the sum of the gap, $\epsilon$, and the total geometric length, $d_0$ (also normalised here by $a$).
For point charge/dipole interactions, $f(\epsilon) = (d_0 + \epsilon)^{-n}$, where $n$ is a positive number.
However, we are faced with a new challenge when dealing with particles/surfaces in close proximity, 
as the cumulative forces between them deviate from point-like behaviour. Hence, $n$ is not necessarily an integer in these cases.
This is because, in physical systems, the net forces are the result of the cummulation of local dielectric interactions and dispersion forces between the molecular constituents of the particles and intercalant media, and their geometry \citep{van_saarloos_soft_2023-1}.
The immediate consequence of this is reflected in the function $f(\epsilon)$ as a general ambiguity in the definition of the length scale, $d_0$.
In our experimental system with a large planar permanent magnet interacting with a tiny sphere, 
the position of the effective induced dipole is expected to concide with the geometric centre of the sphere due to symmetry, whereby $d_0 = 1$.

To reiterate, the majority of known varying forces typically follow power-laws of the form $f(\epsilon) = (d_0+\epsilon)^{-n}$, 
or $f(\epsilon) = \epsilon^{-n}$. Indeed, our data -- \cref{fig:fig1}(b) and (c) -- illustrate two such cases, where $f(\epsilon)=\text{const.}$, and $f(\epsilon)=(1+\epsilon)^{-0.75}$. 
We are now ready to explore the effects of a varying driving force on the particle trajectory, as opposed to that of a constant force.

\subsection{The trajectory of a magnetically driven sphere} \label{sec:magnetic}
We will begin with the analysis of our experimental system (\cref{fig:fig1}(c)), where $f(\epsilon) = (1+\epsilon)^{-n}$.
Invoking \cref{eq:norm_dyn}, the dynamics of the magnetically driven sphere can be written as
\begin{equation}
    \frac{\dot{\epsilon}}{\epsilon} + \frac{1}{(1 + \epsilon)^{n}} = 0.
    \label{eq:mag_dyn}
\end{equation}
As with the drag force, in \cref{sec:drag}, we can seek the governing equation at the limit that $\epsilon \rightarrow 0$, 
whereby \cref{eq:mag_dyn} becomes
\begin{equation}
    \dot{\epsilon}+ \epsilon -  n \ \epsilon^2 = 0.
    \label{eq:bernoulli}
\end{equation}
This equation is an instance of the Bernoulli differential equation, the solution to which is 
\begin{equation}
    \epsilon(\tau) = \frac{\epsilon_0}{ n \epsilon_0 + \left(1- n {\epsilon_0}  \right) e^{\tau - \tau_0} }.
    \label{eq:mag_sol}
\end{equation}
Consistent with  gravity-driven case (\cref{eq:grav_sol}) the initial condition is $\epsilon_0 = \epsilon(\tau_0)$, and the non-dimensional time is $\tau = t/T$, where $T = {6 \pi \mu a^2}/{F_0}$. 
We had seen that, for the case of gravity-driven settling, $F_0 = F_g$, and thence the characteristic time is $T \sim a^{-1}$. 
For power-law interactions, however, the qualitative and quantitative dependence on particle size $a$ differs.
First, we note that in the present case, the magnetic attraction in full dimensional form will be $F = M / (a + \delta)^n$, where $M = F_0 a^n$, and $\delta = a \epsilon$, as previously defined (see \cref{fig:fig1}(a)).
The classical approach to calculating induced dipole interactions in a sphere placed in a uniform external magnetic field suggests that 
the coefficient, $M$, scales with the volume, $V\sim a^3$, of the sphere (for example, §9.06 of \citep{smythe_static_1989-1}). 
Incorporating this, we get $F_0 = m/a^{n-3}$ for $M = m a^3$.
Consequently, the time scale over which magnet-driven settling occurs scales as $T \sim a^{(1-n)}$.                                                       

\begin{figure}
    \centerline{\includegraphics[width=1\linewidth]{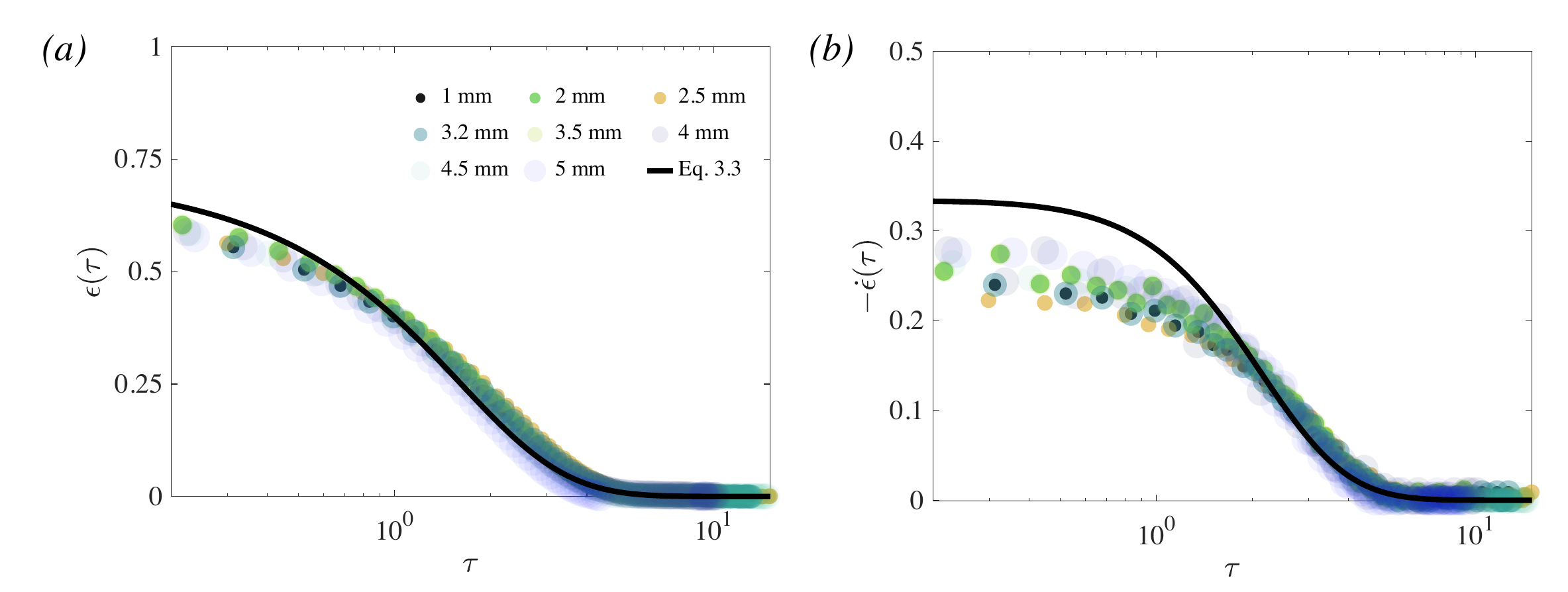}}
    \caption{(a) End trajectories of spheres of different radii, $a$, approaching the surface under magnetic forcing, a varying force of the form $F = F_0/(1+\epsilon)^{n}$. 
    The black line denotes the analytical solution \cref{eq:mag_sol} for the limit $\epsilon \rightarrow 0$.
    Here $F_0 = m/a^{n-3}$ for $m = 10^{5.8}$, and $n = 0.75$, from direct measurements.
    (b) The instantaneous velocity of the sphere, $\dot{\epsilon}$, peaks at at $\epsilon = 1/2n$ 
    following which the trajectory reduces to an exponential decay in nondimensional time $\tau$. 
    Compared to a constant driving force of similar magnitude, the varying force drives the particle across the thin lubrication layer faster by $\Delta \tau \approx \ln{[n \epsilon_0/(1 - n \epsilon_0)]}$.}
    \label{fig:mag_tracks}
\end{figure}

To comprehend what the solution, \cref{eq:mag_sol}, entails intuitively, we can revisit \cref{eq:bernoulli} and impose the limit $\epsilon \rightarrow 0$.
Ignoring the $\mathcal{O}(\epsilon^2)$ term, we are immediately left with the equation $\dot{\epsilon}+ \epsilon = 0$, which is the same as \cref{eq:grav_eqn}.
This implies that the final trajectory exhibits the exponential decay characteristic of a constant driving force (in our experiments, gravity),
sufficiently close to the surface, but over a different time scale $T = {6 \pi \mu a^2}/{F_0}$.
We observe exactly this behaviour in the trajectory data shown in \cref{fig:mag_tracks}(a). 
The effect of a variable force that follows $f(\epsilon) = (1+\epsilon)^{-n}$ manifests as a quick descend towards the surface, followed by an exponential fall over the rest of the path.
In in the magnet-driven trajecory laid out in \cref{fig:fig1}(c),
the sphere traverses the distance from $\epsilon = 0.1$ to $\epsilon = 0.01$ in 0.0832s, as opposed to 10.06s for the gravity-driven sphere in \cref{fig:fig1}(b).
The solutions to \cref{eq:mag_sol} for other values of $n$ are plotted in \cref{fig:fam_sol}(a).

In order to discern where the transition to `constant-force' behaviour sets in, we can take a look at the nature of $\dot{\epsilon}$, the rate at which the gap closes. 
From the plots of $\epsilon(\tau)$ and $\dot{\epsilon}(\tau)$ in \cref{fig:mag_tracks}, notice that the rate of descend increases over time, 
and briefly achieves peak velocity, before dropping exponentially towards the end of its trajectory.
Seeking the local maxima of $\dot{\epsilon}$, we find that the critical point is $\epsilon_c = 1/2n$ at nondimensional time $\tau_c = \tau_0 + \ln{[n \epsilon_0/(1 - n \epsilon_0)]}$.
Beyond this critical point, the exponential solution begins to dominate the kinematics of the sphere.
Thus it can be concluded that the sphere driven by such a force, $F \sim (1+\epsilon)^{-n}$, traverses the thin gap beginning at $\epsilon_0(\tau_0)$ 
faster than when settling under a constant force of the same magnitude, $F_0$, by a duration of approximately $\Delta \tau \approx \ln{[n \epsilon_0/(1 - n \epsilon_0)]} $.

\subsection{Determining $f(\epsilon)$ from kinematic data} \label{sec:estimate_forces}
Before moving on with the study of the other class of variable forces, namely, $(f(\epsilon) =  \epsilon^{-n})$, 
we have one major consequence of the above analysis to elaborate upon.
Revisiting \cref{fig:mag_tracks}, the remarkable correspondence between theory and experiments gives us the opportunity to estimate the driving force from the kinematic data.
Since in this case $F = F_0 / (1 + \epsilon)^{n}$, this involves determining the values of $F_0$ and $n$.
To validate the values extracted from the data, we compare the estimates with the measured force function, produced in \cref{fig:mag_force} in \cref{appB}.
We find that the forces fit best (by least squares) with $n = 0.75$ and $m = 10^{5.8}$ for $F_0 = m/a^{n-3}$.
These values have been used in calculating the characteristic kinematic time scale, $T$, which has been used in normalising the time axis in \cref{fig:mag_tracks}.

This method could be of special relevance to studying systems where \textit{in situ} force measurements are impossible to realise, such as in biological samples.
In order to implement this, the spatial resolution of data capture must be adequate to probe the range $\epsilon \ll 1$ with a temporal resolution $\sim T$.
For instance, for a vesicle ($a \sim 1 \mathrm{\mu m}$) driven by cytokinetic actin-myosin networks with a force of magnitude $\sim 100 \mathrm{\mu N}$ \citep{phillips_physical_2013},
this corresponds to an image resolution of $\sim 0.01 \mathrm{\mu m}$ in time steps of $ \sim 2 \mathrm{\mu s}$.
Microscopy being the primary non-invasive tool available to us, data of such resolution will allow extracting information of $n, F_0$ from kinematic data.

\section{To snap or not to snap} \label{sec:snap}

\begin{figure}

    \centerline{\includegraphics[width=1\linewidth]{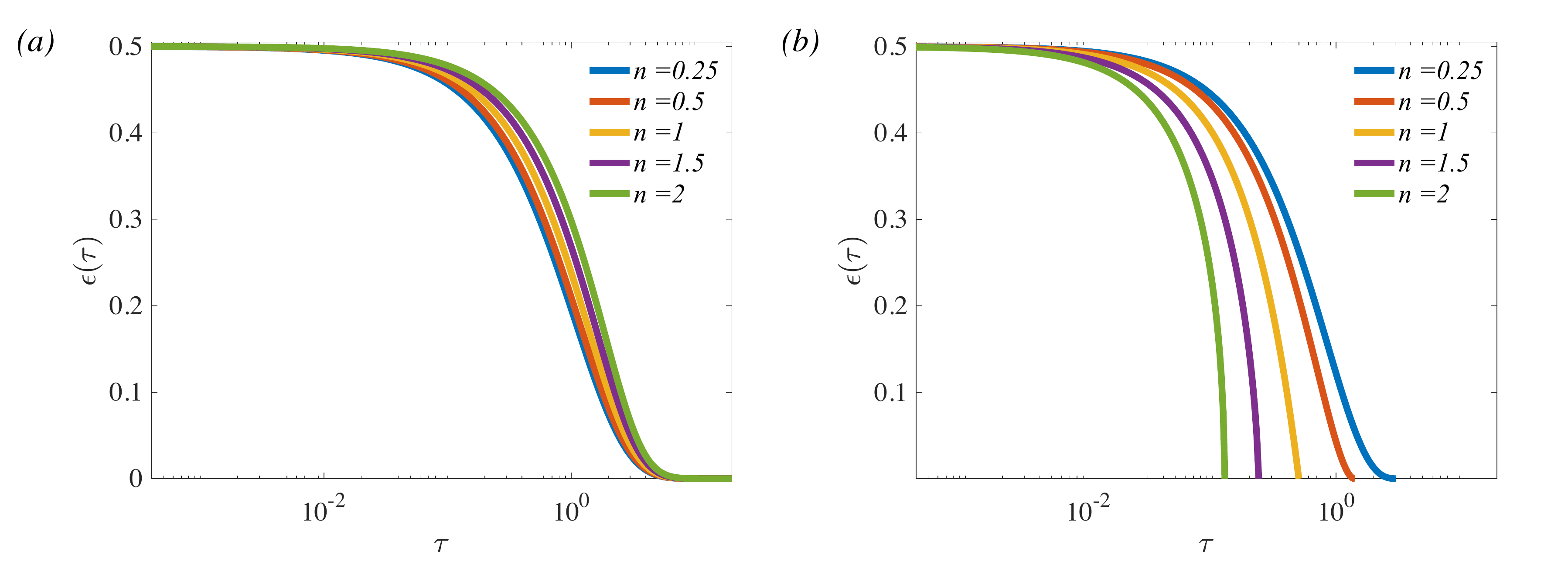} }

    \caption{Family of particle trajectories approaching a flat surface in response to spatially varying driving forces. The plots indicate the effect of the exponent, $n$ (see \cref{force_table}), 
    for forces of the same magnitude $F_0$, with the same initial condition $\epsilon(0) = \epsilon_0 = 0$. 
    (a) Solutions of \cref{eq:mag_dyn} where a spherical particle is driven by a force of the form $f(\epsilon) = (1 + \epsilon)^{-n} $, such as by magnetic forcing realised in our experiments.
    (b) Solutions to \cref{eq:vdw} which governs particle kinemtics in response to $f(\epsilon) = \epsilon^{-n} $ for different values of $n$, (see \cref{force_table}). 
    Note that these trajectories achieve contant, i.e., $\epsilon = 0$ in finite time, at $\tau = \epsilon_0^n /n$.
    Furthermore, the particle is seen to briefly decelerate prior to contact when $n<1$, while it `snaps' onto the surface when $n>1$. }
    \label{fig:fam_sol}
\end{figure}

In the preceding sections, we explored the possibility how a spatially varying force may be the key method to overcome the 
thin film drag that varies inversely with the gap, $\epsilon$, in order for the surfaces to achieve contact in finite time.
However, we saw that forces that vary as $f(\epsilon) = (1+\epsilon)^{-n}$ result in interesting kinematics of the sphere 
as it approaches the flat plane.
As the above analysis revealed, while the trajectories of magnet-driven spheres in \cref{fig:mag_tracks} approach the plane at faster rates,
they transition to an exponential path towards the end of their paths.
This leaves a major question in our investigation unanswered: can a particle traverse the intercalant fluid at the thin gap limit in finite time?

To resolve this we can proceed now with the second major class of $f(\epsilon)$, namely forces that vary as $\epsilon^{-n}$. 
Several familiar, well-formulated forces, such as van der Waals interactions, London dispersion forces, and Stefan adhesion follow such a power-law, the respective 
values of $n$ for which are consolidated in \cref{force_table}.
Notably, this class of interactions are devoid of additional (geometric) length scales such as $d_0$, unlike the magnetic force contrived in our experimental setup.
We can begin by rewriting \cref{eq:norm_dyn} for this case as
\begin{equation}
    \frac{\dot \epsilon}{\epsilon} + \frac{1}{\epsilon^{n}} = 0.
    \label{eq:vdw}
\end{equation}
In constrast to the corresponding \cref{eq:mag_dyn}, we can directly integrate the above equation to yield
\begin{equation}
    \epsilon(\tau) =\left[ \epsilon_0^n  -n \left( \tau - \tau_0 \right) \right]^{\frac{1}{n}}.
    \label{eq:vdw_sol}
\end{equation}
It is worthwhile to visualise the above solution for an expository set of values of $n$ to highlight two of its notable features, depicted in \cref{fig:fam_sol}(b).  
The first remarkable consequence of such a trajectory is the scale-free evolution of the gap in time. 
Thus, the particle is guaranteed to achieve contact in a finite time, crossing the thin film barrier from $\epsilon_0(\tau_0)$ to $\epsilon = 0$ within the duration, $\Delta \tau = \epsilon_0^n /n$.
Secondly, we can focus on the slope of the curves at touch down. 
From \cref{eq:vdw}, we readily have $\dot{\epsilon} = - \epsilon^{1-n}$. 
Clearly, for $n = 1$, the trajectory falls at a constant rate of $-1$.
This is a peculiar case where both the driving force and fluid drag scales inversely with $\epsilon$, and balances one another perfectly.
It is interesting to note that $n=1$ forces have not been reported, to the best of our knowledge, 
and hence do not appear on \cref{force_table}, which is considated with examples from prominent literature.
For $n<1$, typically observed between macroscale particles or clusters of point dipoles,
$\dot{\epsilon}$ can be seen to diminish as the gap closes, decelerating briefly before contact.
In contrast, when $n>1$, i.e., for short range forces such as van der Waals, London, and Stefan interactions, 
the magnitude of $\dot{\epsilon}$ increases prior to contact, such that the particle may be considered to `snap' onto the planar surface.

\section{Discussion and Conclusion} \label{sec:conclusions}

Recapitulating the conclusions from the preceding sections, we recognise that the two major classes of spatially varying forces, $F = F_0 f(\epsilon)$, 
where $f(\epsilon) = (1+\epsilon)^{-n}$ and $f(\epsilon) = \epsilon^{-n}$ result in strikingly different kinematics of a particle approaching a plane.
Both of these generalised power-law forces aid the particle in overcoming the thin film drag to achieve contact to different extends.
We saw that while forces that vary as $(1+\epsilon)^{-n}$ eventually results in an exponentially devolving trajectory akin to that due to a constant force, 
it delays the onset of the exponential behaviour to much smaller gap heights.
On the other hand, forces of the form $F \sim \epsilon^{-n}$ definitively achieves contact in finite time.
We ought to remember that  both types of forces coexist in physical systems.
Revisiting \cref{force_table}, we see that forces that follow $f(\epsilon) = \epsilon^{-n}$ are typically short range dispersion forces, 
which become dominant only at very close proximity between interacting surfaces or particles. 
Hence, we can presume that regardless of the form and magnitude of the driving force at finite separation, 
towards the end of its trajectory, a particle may be brought to contact by a short range dispersion force, 
i.e., particles on paths plotted in \cref{fig:fam_sol}(a) could transition to trajectories in \cref{fig:fam_sol}(b) prior to contact.

It is interesting to note that one may come across force laws that do not immediately follow from standard interaction forces.
Namely, following the standard assumption that interaction energies, $\mathit{U}$, are additive \citep{israelachvili_intermolecular_2011},
one may encounter cases where particles exhibit the counterintuitive law, $f(\epsilon) = 1/(1-\epsilon)$, 
which follows from the infinite geometric series $\sum_{k=0}^{\infty} \epsilon^k = 1/(1 - \epsilon)$ for $|\epsilon| < 1$, 
consistent with our range of $\epsilon$, when a particle experiences a combination of numerous repulsive, dispersion forces of similar magnitude.

In this study, we have shown how a spatially varying driving force, $F = F_0 f(\epsilon)$, can effectuate inter-particle/ particle-surface contact, 
overcoming the thin film barrier in finite time.
An alternative approach would be to modify the effect fluid drag, $F_D$, at the thin film limit, so that the obstructive force no longer scales as $\epsilon^{-1}$,
such as by modifying the surface properties of the particle/ wall to introduces surface interactions that evolve over time scales comparable to $T$, relevant to the kinematics detailed in this study.
This may be achieved via exploiting tunable chemical reaction rates \citep{yariv_wall-induced_2016} or active polymer grafts \citep{xu_dynamic_2023}.

A major consequence of our analysis is the possibility of estimating the driving force, $F$, from kinematic data, as elucidated in \cref{sec:estimate_forces}.
Conversely, a better deterministic understanding of varying forces afforded by our analysis may be exploited to improve processes that are 
reliant on surface interactions, given the advent of self assembly as a major tool in fabrication processes.
Since this study was motivated by our experiments, where a varying force was realised using steel spheres interacting with a permanent magnet,
our focus has been limited to rigid, smooth particles.
Nonetheless, our analysis may, in principle, be extended to systems of other geometries \citep{stone_lubrication_2005-3}, where the same interplay of forces can be observed between surfaces in close proximity.
In numerous scenarios, transient deformations of interacting surfaces can result in more complex dynamics, which we leave for future investigations.

\newpage

\appendix
\section{}\label{appA}

\subsection*{Experimental setup and data capture}

The experimental system briefly described in \cref{sec:intro} and depicted schematically in \cref{fig:fig1}(a),
comprises a transparent, cuboidal acrylic box of side 10 cm, bearing a high viscosity silicone oil (Sigma-Aldrich 200) of 
kinematic viscosity = $60,000$ cSt (at 25\textdegree C) and specific gravity of 0.98.
Steel ball bearings (WJMY Precision Chrome Steel Bearings) of eight different diameters $2a = 1, 2, 2.5, 3.2, 3.5, 4, 4.5, 5$ mm, were used.
The density of the steel alloy, $\rho_s = 7780\ \mathrm{kgm^{-3}}$. 
The spheres were cleaned and wetted using the same oil prior to carefully dropping them into the fluid bath along the centre of the container.

\begin{figure}
   \centerline{ \includegraphics[width=1\linewidth]{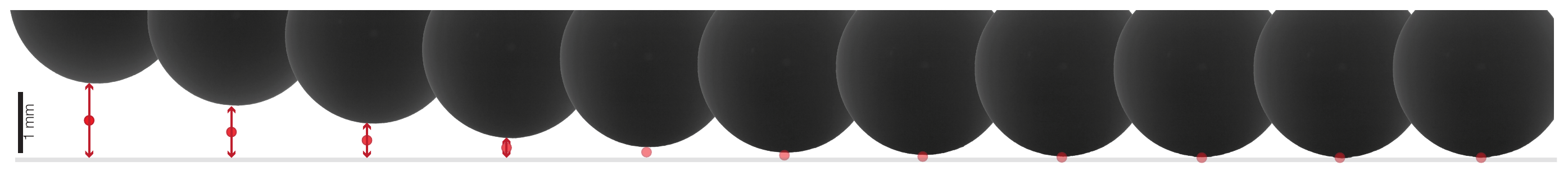} }
    \caption{A montage of selected frames from the end trajectory of a $a = 1.25$ mm sphere settling under magnetic attraction captured using a high speed camera with a spatial resolution of $2.23 \mathrm{\mu m}$.
    The time interval between successive frames is $0.953 \mathrm{ms}$.
    The red arrows denote the measured quantity, the gap, $\delta$. }
    \label{fig:kymo}
\end{figure}

The trajectories of each sphere is captured using a high speed camera (Chronos 1.4, Kron Technologies).
As each sphere translates at a different rate, and traverses the height of the frame over different durations,
the maximum possible framerate was chosen for each sphere. 
Temporal resolution therefore ranges from 0.9 to 5.6 ms across the experiments. 
The spatial resolution of image data falls in the range 2.23 - 2.32 $\mu \mathrm{m/pixel}$. 
For each radius, the experiment was repeated at least four times each for gravity- and magnet-driven cases. 
From the video data, the trajectory was extracted by tracking a single point at the border of the sphere which is seen as a circle in the backlit frames,
as shown in \cref{fig:kymo}. 
The calibrated data from the multiple repetitions were binned and compiled.
For magnet-driven settling experiments, modifications were made only to the settling tank, which are described below.

\subsection*{The magnetic system}
The concept behind the implementation of the magnetic, spatially varying force, in our experiments is rather straightforward.
Placing a permanent magnet of \textit{infinite} extents beneath the fluid reservoir in the sedimentation apparatus described in the preceding section 
would overlay a uniformly directed magnetic field with its magnitude decaying from its peak strength at the base ($z=0$) towards the top of the tank.
However, since all finite magnets exhibit edge effects we modify the system further using what we label as a `diffuser lens' for the magnetic field lines.
The magnet that we use at the base of the reservoir is a cylindrical neodymium (N45) magnet of diameter 70 mm and height 35 mm (Supermagnete, Germany).
From the magnetic field data reported by \citet{abelmann_permanent_2024-1} (see Figs 3 and 4 therein), for a magnet of the same dimesions, make and source,
we identify a region of uniform, axisymmetric field in the central region of the magnet.
However, at the limit of thin gaps that we investigate in this work, both the magnetic field and its gradients are strong enough to induce transverse forces on the sphere, 
given the axisymmetry of the magnetic field.

To improve the uniformity in direction of the magnetic field, and consequently the force experienced the sphere to be unidirectional along $z$,
we draw inspiration from the concept of  Halbach arrays \citep{jae-seok_choi_design_2008}. 
Noticing the stracked arrangement of slender grains on a cuboidal ferrite magnet of side 25 mm, we place it within the tank, 
to act as a diffuser/ lens for the magnetic field. 
In order to avoid skewing of the magnetic field due to the presence of the ferrite block, 
we first allow the block magnet to come to its rest position, and perturb it over several cycles, till it consistently returns to the equilibrium position.
The uniformity of the diffused field is noticeable at once from particle trajectories, which remain staunchly vertical.
For the magnetic experiments, the metallic sphere settles onto the top surface of the submerged ferrite block,
which is defined as $z = 0$, rather than the base of the reservoir.

\section{}\label{appB}
\begin{figure}
    \centerline{ \includegraphics[width=0.7\linewidth]{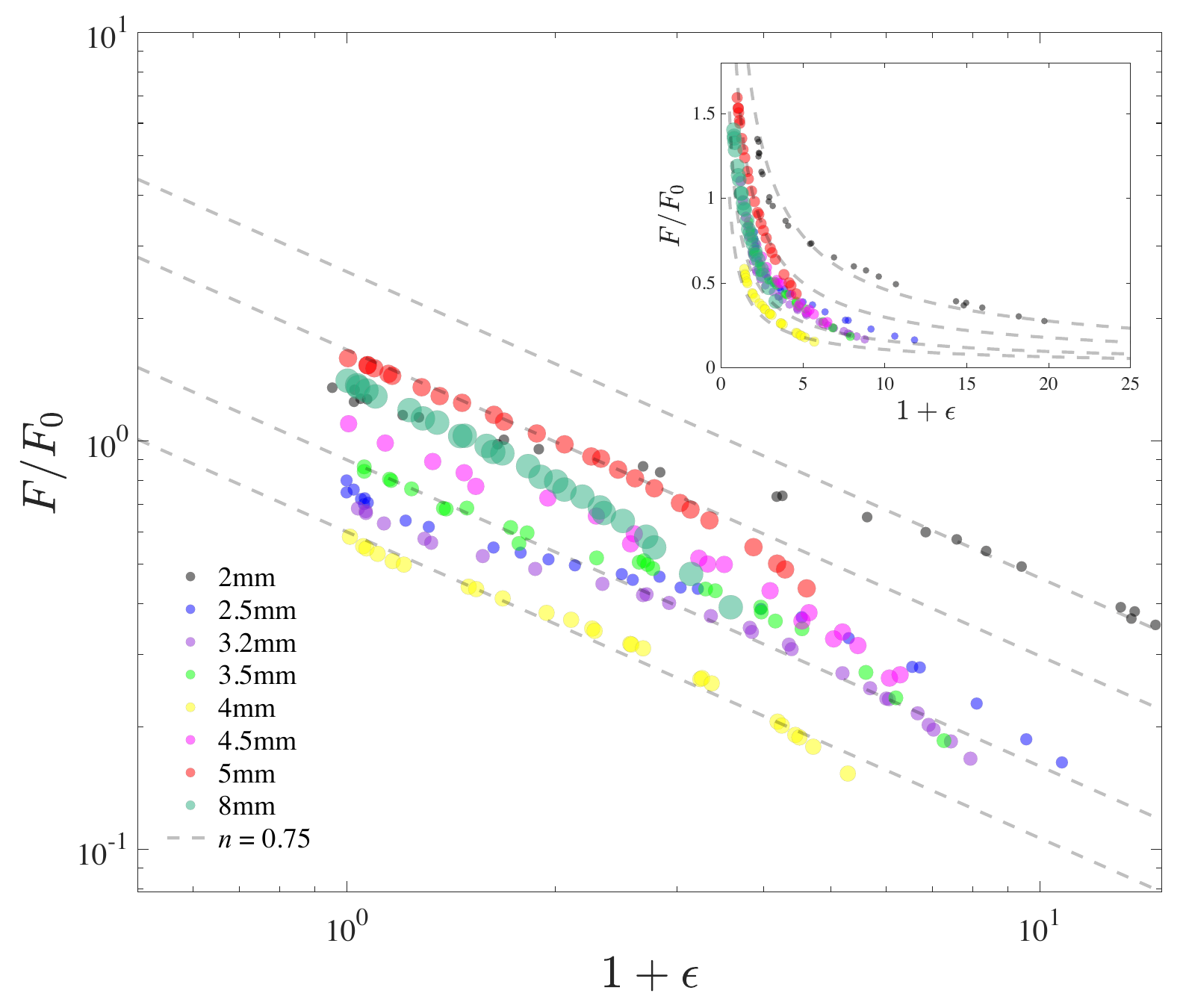} }
    \caption{Direct force measurements of the magnetic attraction induced in the metallic spheres are presented as ${F}/{F_0} -  (1 + \epsilon)$.
    The force follows a power-law of the form $f(\epsilon) = (1+\epsilon)^{-n}$ which is plotted as the dashed line for the best fit value of $n = 0.75$.}
    \label{fig:mag_force}
\end{figure}
In order to validate the theoretical model describing the magnet-driven trajectory, namely, the exponent $n$ and 
$F_0$ which manifests as the denominator in the characteristic time scale, $T$,
we make direct measurements of the induced magnetic attraction.
A selection of spheres is attached to a pedestal extending from the tip of a cantilever load cell (TAL221, HT Sensor Technology Co.; HX711 ADC, M5Stack).
The purpose of the pedestal is to isolate any magnetic attraction on the metallic load cell tip may experience directly,
and is therefore made of a polymer (PETG).

The force data is recorded at different values of gap separation, $\delta$, along the centreline of the reservoir, 
retracing the same path along which the spheres travel, to determine $F = F_0 f(\epsilon)$ relevant to our experiments.
Gap thickness is found using the same video camera. 
The force data is produced in \cref{fig:mag_force}, where the measured force $F$ is normalised by $F_0 = m/a^{n-3}$.
The dashed line is the equation $F/F_0 = 1/(1+\epsilon)^{n}$ for $m = 10^{5.8}$ and $n = 0.75$.

Unsurprisingly, the force data does not collapse onto the theory line in \cref{fig:mag_force} for all $a$,
for reasons briefly mentioned in \cref{sec:varying}.
In the far field, it is immediately apparent that this centre of magnetic force coincides with the geometric center of the sphere, whereby $d_0 = a$,
as we have assumed in our work.
However, when the sphere is close to the surface, the effective dipole may shift from the centre of the sphere, closer to the base, due to the large local gradient in the magnetic field.
We observe that this shift is pronounced only in spheres of larger radii. 
Therefore, our analysis is limited to relatively smaller values of $a$, where the induced dipole remains coincident with the geometric centre, i.e., $d_0 \approx 1$, 
keeping with the various examples that has motivated this study.

\bigskip
\noindent \textbf{Declaration of Interests.} The authors report no conflict of interest.

\noindent \textbf{Author ORCIDs.} \\
\orcidlink{0000-0003-3430-1349}John Sebastian \url{https://orcid.org/0000-0003-3430-1349} \newline
\orcidlink{0009-0003-9253-5968}Alexander Lukinych Schødt \url{https://orcid.org/0009-0003-9253-5968}\newline
\orcidlink{0000-0003-0787-5283}Kaare Hartvig Jensen \url{https://orcid.org/0000-0003-0787-5283}

\bibliographystyle{jfm}
\bibliography{galileo}
\end{document}